# Photons in - numbers out: Perspectives in quantitative fluorescence microscopy for *in situ* protein counting

K. S. Grußmayer[1, #], K. Yserentant[2, #], D.-P. Herten[2,*]

[1] *Laboratory of Nanoscale Biology, Faculté des Sciences et Techniques de l'Ingénieur, Ecole Polytechnique Fédérale de Lausanne, Station 17, 1015 Lausanne, Switzerland*
[2] *Institute for Physical Chemistry, Heidelberg University, Im Neuenheimer Feld 229, D-69120 Heidelberg, Germany*

[*]Corresponding author: dirk-peter.herten@urz.uni-hd.de

[#]These authors contributed equally to this work.

## Abstract

The full understanding of cellular functions requires information about protein numbers for various biomolecular assemblies and their dynamics, which can be partly accessed by super-resolution fluorescence microscopy. Yet, many protein assemblies and cellular structures remain below the accessible resolution on the order of tens of nanometers thereby evading direct observation of processes, like self-association or oligomerization, that are crucial for many cellular functions. Over the recent years, several approaches have been developed addressing concentrations and copy numbers of biomolecules in cellular samples for specific applications. This has been achieved by new labeling strategies and improved sample preparation as well as advancements in super-resolution and single-molecule fluorescence microscopy. So far, none of the methods has reached a level of general and versatile usability due to individual advantages and limitations. In this article, important requirements of an ideal quantitative microscopy approach of general usability are outlined and discussed in the context of existing methods including sample preparation and labeling quality which are essential for the robustness and reliability of the methods and future applications in cell biology.

Keywords: single-molecule fluorescence microscopy; protein counting; fluorescence labeling; quantitative microscopy; fluorescence standards

# Introduction

Over the past decades, the technical progress in fluorescence microscopy enabled the study of structure-function-relationships in living cells at nanometer resolution with single-molecule sensitivity[1]. Despite these advances in optical microscopy and fluorescent probe design many important protein assemblies elude the collection of quantitative data, since they are smaller than the current limit of super-resolution. Important dynamic processes, such as self-association and oligomerization, are prominent examples, crucial for the functioning of many proteins[2–5]. For example, Nanoscale clustering of membrane proteins emerged as a common feature, possibly initiating and amplifying signal transduction across the membrane, e.g. in the immune response of T cells[6]. Direct access to copy numbers of participating proteins is a prerequisite for a quantitative technique to study these processes. This motivates the search for alternative, widely applicable methods allowing quantitative assessment of oligomeric structures and the kinetics of their formation, transformation, and disassembly.

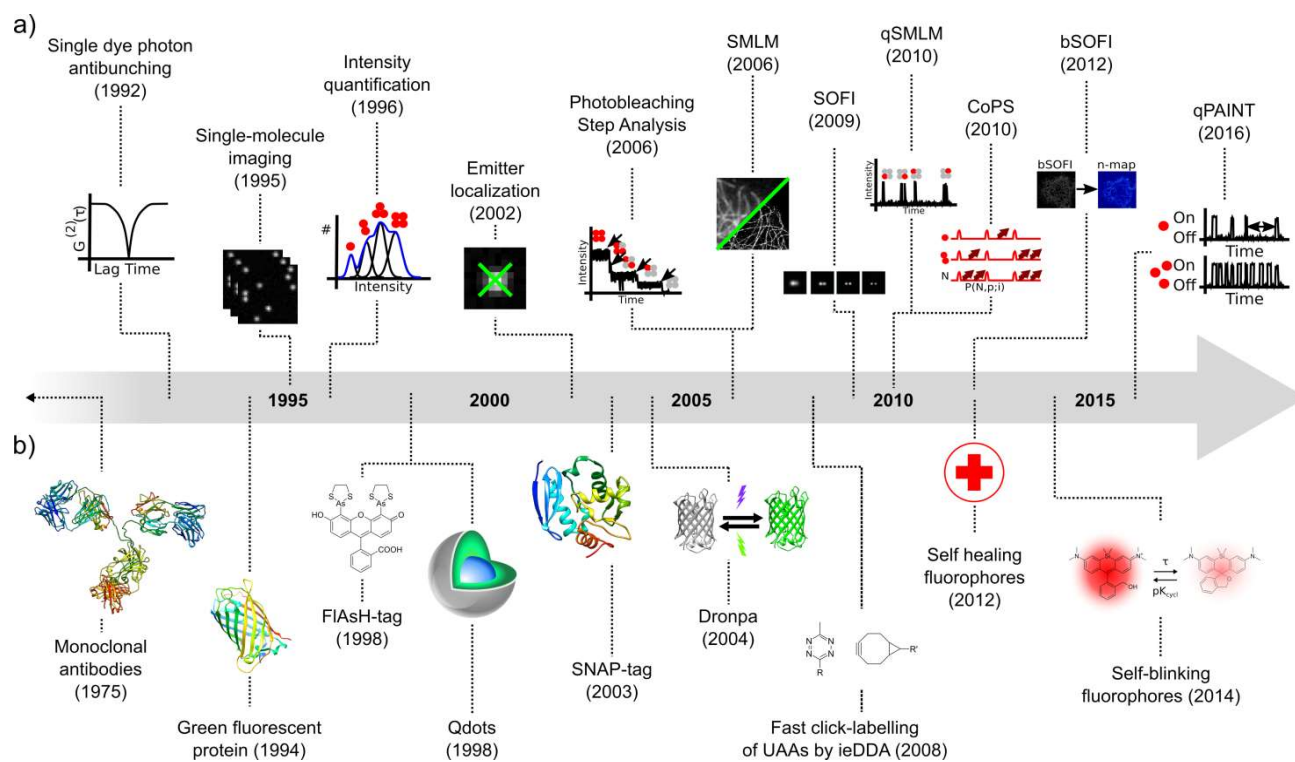

**Figure 1: Major developments in fluorescence microscopy (a) and fluorescent labeling (b) enable quantitative approaches to fluorescence microscopy.** a) Antibunching measurements on single dyes paved the way for counting by photon statistics (CoPS) analysis. Development of instrumentation for single-emitter imaging allowed for protein counting by photobleaching step analysis and for the development of (balanced) super-resolution optical fluctuation imaging((b)SOFI). Single-emitter imaging in combination with emitter localization enabled development of (quantitative) single-molecule localization microscopy ((q)SMLM) and quantitative point-accumulation imaging in nanoscale topology (qPAINT). b) Monoclonal antibodies were among the first labels that allowed for molecule-specific labeling. Fluorescent proteins such as green fluorescent protein (GFP) and photo-switchable fluorescent proteins such as Dronpa allow for quantitative fluorescence imaging without addition of external dyes. FlAsH- and SNAP-tag, as well as unnatural amino acids (UAAs) enable conjugation of organic fluorophores to proteins in live samples. The development

of quantum dots (Qdots) allowed for applications where exceptionally high brightness an photostability is required. Recent developments include advanced organic fluorophores with additional features such as self-healing and self-blinking fluorophores. Protein structures in b) were rendered using Chimera[7] based on publicly available data deposited in PDBe. Accession numbers: 2ie2, 2Y0G, 1IGT, 3KZZ.

In recent years, several fluorescence-microscopy-based approaches for measuring protein copy numbers emerged, e.g. based on emission intensity, photon or blinking statistics, or distinguishable bleach-steps[8–12] (fig. 1a). This development was supported by different labeling approaches developed during the same time (fig. 1b). In the following, we will confine ourselves on fluorescent proteins (FPs), protein- and peptide-tags (PP-tag) and affinity-based labeling as these are the most common and well-established classes of fluorescence labeling methods. In terms of fluorescence-microscopy-based methods for protein counting, we will focus on methods based on fluorescence intensity, single-molecule localization microscopy (SMLM), and photo-bleaching steps which have already frequently been used, as well as balanced super-resolution optical fluctuation imaging (bSOFI), quantitative point-accumulation imaging in nanoscale topology (qPAINT) and counting by photon statistics (CoPS) which have emerged more recently showing interesting perspectives in future developments. Thus far, these protein counting methods were used to solve specific problems because they bear individual advantages and limitations. Motivated by the lack of a universal quantitative microscopy technique, this article addresses the question of what an ideal method would require. These demands then set the context for a comparison of existing methods and a discussion of their future developments. First, we will outline basic requirements of an ideal method from the perspective of potential users. As any ideal technique will count fluorophores rather than target proteins we will first elaborate on approaches for fluorescence labeling and sample preparation in general which can have a strong influence on the apparent label number. Then, existing approaches to quantitative microscopy shall be briefly discussed in this context. Finally, the comparison of the current quantification methods and their workflows in conclusion leads to a perspective for future developments.

**Requirements for an ideal method**

From a user's perspective any quantitative fluorescence microscopy approach must be compatible with biological samples, ideally applicable to living cells, and non-invasive toward relevant biological functions. This requires the method to work within a crowded cellular environment. Although fluorescence microscopy is said to be non- or minimally invasive, the required fluorescent labeling might alter protein function making additional control experiments necessary. Lately, photo-toxicity reached a higher attention as various artifacts in cellular structures have been observed at high irradiation intensities used in super-resolution microscopy[13,14]. This photo-damage, which is especially severe at low wavelengths, is showing a clear demand for low-intensity fluorescence microscopy when dealing with live-cell samples and motivates research into e.g. alternative (photo-)switching mechanisms such as primed conversion to avoid exposure to UV light [15,16].

Aside of a high labeling specificity, a high labeling efficiency needs to be reached and verified, as deviations will distort copy number estimates. Consequently, one should also account for the question of unlabeled endogenous protein in the samples. Although various ways for fluorescent labeling exist enabling specific imaging with high contrast, any method will have to deal with at least a low fluorescent background and potentially optically overlapping structures. The latter will result in a broadening of the distribution of protein-number estimates thereby increasing the measurement error. Ideally, the target structures in question should be displayed in a micrograph were image contrast is directly reflecting the number estimates of protein copies present to allow direct correlation with cellular structures. Additionally, the number estimates should be absolute without requirement of additional calibration measurements to enable online display offering high experimental control. Generally, the precision and bias of the quantitative imaging modality need to be well characterized, enabling robust measurements with reliable results. Finally, an ideal method would be non-destructive and allow time-resolved data acquisition to determine the kinetics of structure formation and disassembly.

**Labeling approaches for protein counting**

Selective labeling with fluorophores enables observing proteins of interest (POIs) by fluorescence microscopy against a vast background of other proteins and small molecules within cells in real time and with single molecule sensitivity. However, achieving specific labeling with high quality fluorophores and low non-specific background can pose a demanding challenge. If target molecules are to be counted with the help of fluorescence, the requirements are even further increased, and a detailed characterization of the labeling quality becomes crucial.

Fluorescent labels in their simplest implementation can be regarded as bi-functional molecules with an anchoring moiety to attach the label to its target and a fluorophore as reporter. The most commonly employed fluorophores for biological imaging are fluorescent proteins, fluorescent organic dyes and fluorescent nanoparticles such as quantum or carbon dots (Qdots, Cdots) [17,18]. While photo-physical fluorophore properties are often of high importance for the counting approach employed, the anchoring mechanism is crucial when it comes to characterizing labeling quality. Depending on the nature of the label, the anchoring moiety may require external factors for attachment to target molecules and/or modification of targets, i.e. by genetic alterations of the host organism.

In principle, labeling quality depends on both, labeling efficiency and specificity. labeling efficiency is typically reported as the degree of labeling (DOL) that is the ratio of fluorescent labels to target molecules. To evaluate the labeling specificity, the off-target label deposition needs to be characterized quantitatively in terms of off-target label density and spatial heterogeneities. Depending on the sample, it may also be required to characterize the contributions of autofluorescence to the measured signal. It is important to note that labeling efficiency and off-target labeling characterizations should be performed under identical labeling conditions used for protein copy number quantification. While the DOL as an average value over an unknown population can be used by any quantitative method to account for a bias in the determined protein copy numbers it cannot distinguish random variations of the oligomerization from those in the labeling stoichiometry. This problem can only be addressed by measuring the label number distribution (LND) using one of the methods discussed below[19].

**Labeling with fluorescent proteins**

Fluorescent proteins (FPs) are inarguably the most widely used fluorescent markers in biology, especially when it comes to live-cell and *in vivo* imaging. Fluorescent proteins were first isolated from cnidarians in the 1960s and 1970s and adapted as marker for biological fluorescence imaging in the 1990s[20–22]. Today, a wide range of fluorescent proteins with different spectral properties based on wild type FPs from several different species is available and comparative studies covering a large number of FPs allow choosing optimal FPs for a given experiment[23–25].

Due to their ease of use, fluorescent proteins are frequently chosen as labels for intensity-based protein copy number estimation and photo-bleaching step analysis as discussed in section 3.2 [11,26]. Based on this line-up of FPs, many photo-controllable fluorescent proteins were developed that can be reversibly switched or irreversibly converted between multiple spectral or emissive states[27–29]. Photo-controllable FPs are mainly being used as labels for super-resolution microscopy approaches such as PALM[30], SOFI[31], or RESOLFT[32] and can also be used for estimating protein copy numbers based on blinking frequencies (section 3.3).

FPs are conjugated to their target via stable or transient expression of a target-FP fusion protein or via introduction of the FP coding sequence into the host organism's genome using gene editing techniques such as CRISPR/Cas9[33]. In both cases, it must be ensured that the conjugation to an FP does not interfere with function and localization of the POI[34,35]. In the context of quantifying protein copy numbers, it is also important to ensure that FP introduction does not affect the stoichiometry of formed complexes. Different systems to fine-tune the expression strength of transgenes, such as variable and inducible promoters are available[36,37]. If endogenous, non-labeled target protein is present at the same time as POI-FP fusions, the ratio of genetically modified to endogenous protein has to be determined e.g. by quantitative western blotting or quantitative mass spectrometry[38]. Alternatively, endogenous protein expression can be suppressed using RNA interference[39]. Both, artifacts due to over-expression of FP fusion proteins, and complex dilution due to expression of unlabeled target proteins can be circumvented using gene editing to generate knock-in FP fusions[40]. Any quantitative approach also must consider that some fluorescent proteins form oligomers, typically dimers or tetramers, in their natural form[41] and, depending on the subcellular localization of the POI-FP fusion, disulfide bond formation may lead to oligomerization as was shown for example for EBFP2[42].

Fluorescent proteins as labels don't require additional labeling steps. However, protein folding as well as chromophore maturation efficiency must be considered to avoid under-counting of target proteins[43]. For EGFP, variable folding efficiencies ranging from 65 to almost 100 % were reported[44,45]. For red-shifted FPs, like mCherry and mScarlet, reported folding efficiencies in mammalian cells were considerably lower (around 40 %)[46].

Different approaches to measure the effective degree of labeling with FPs have been described in the literature. Foo et al. used fluorescence cross correlation spectroscopy to determine the FP maturation efficiency and hence the apparent degree of labeling for EGFP-mCherry tandems[44]. By measuring the molecular brightness of FP homo-dimers using fluorescence correlation spectroscopy, Dunsing et al. determined the maturation efficiency of a number of FPs under different conditions[46]. A fundamentally different approach is based on measuring the number of FPs for protein complexes with known oligomerization state. Such approaches were demonstrated to work with bacterial enzyme complexes that were expressed as FP fusions in mammalian cells[47], as well as for determining the maturation and photo-activation and photo-conversion efficiencies of photo-controllable FPs using plasma membrane receptors as templates[48].

**Labeling with protein and peptide tags**

Protein and peptide (PP-) tags form a heterogeneous class of labels that are different from fluorescent proteins in that the tags themselves are not fluorescent. Instead, the tag which is usually incorporated into the sample by means of transgene expression binds a fluorescent substrate[49–51]. While labeling with PP-tags is therefore more complex, the use of small organic dyes as substrates comes at the advantages of higher brightness and photo-stability of the fluorophores as well as more flexibility as different substrates can often be used to label the same tag. Protein and peptide tags vary with respect to tag and label size, substrate binding mechanism, substrate specificity and reactivity. While protein tags form isolated domains, peptide tags are smaller in size and, usually, don't fold into functional domains. Unnatural amino acids (UAAs) can be considered the extreme case of a peptide tag with the minimally possible tag size of a single amino acid[52]. Substrate binding of protein and peptide tags can be covalent or non-covalent. While some tags directly recognize fluorophores as ligands, other tags bind a non-fluorescent structural motif that then serves as linker to guide the attached fluorophore to the tag thereby allowing the targeting of different fluorophores to such tags. Enzyme tags are a third class of protein tags which combine the specificity of enzyme-based fluorophore attachment with the small size of peptide tags[50].

FlAsH tag was among the first non-FP for protein labeling within cells[53]. This peptide tag consists of a 12 amino acid targeting sequence including a tetracysteine motif fused to the POI and bisarsenical fluorophores irreversibly binding to the targeting peptide[54]. SNAP-tag is a commonly used protein tag which was developed by mutation of the enzyme $O^6$-alkylguanine DNA alkyltransferase to form a suicide enzyme that covalently binds substrate molecules containing a benzylguanine moiety as linker[55,56]. labeling of the FlAsH tag requires specially developed fluorophores, which have to be chosen based on their ability to bind the corresponding tetracysteine tag and not due to the required photo-physical properties[57]. Self-labeling protein tags such as the SNAP-tag on the other hand allow the use of labels containing a large range of synthetic dyes as reporters since label recognition occurs via structural motifs that are independent from the fluorophore[58,59]. Table 1 provides an overview of some commonly employed protein, peptide and enzyme tags for fluorescence microscopy, all of which are, in principle, suited as labels for protein counting experiments. Further examples are listed in more specialized reviews[51,60,61].

Despite the substantial differences, PP-tags share common properties when it comes to determining the achieved degree of labeling. Since such tags require an additional labeling step with a fluorescent substrate, both genetic fusion of the target protein with the protein or peptide tag and labeling with the fluorescent substrate must be characterized with respect to their efficiencies. In addition, labeling based on PP-tags requires characterization of non-specific substrate deposition during the secondary labeling step to avoid over-counting. For example, substantial differences in off-target label deposition were observed for different SNAP-tag substrates[62]. Since PP-tags require genetic modification of the target, the above discussed implications for genetically encoded labels, like FPs, also need to be considered when choosing PP-tags as labels. While for the majority of PP-tags, no absolute labeling efficiencies are reported so far, few examples for tags with reported DOLs exist (table 1). However, the approaches and conditions under which the DOL has been determined vary greatly and complicate comparisons among substrates and tags. Wilmes et al. measured the DOL for labeling with SNAP-tag using single-particle tracking[63]. Their reported DOL of 0.43±0.05 upon labeling under live-cell conditions differs substantially from previous reports where DOLs of 0.7-0.95 for SNAP-tag were determined using in vitro ensemble measurements with purified proteins[59,64]. Using a third approach where SNAP-tag was labeled after fixation and permeabilization of cells, Finan et al. determined a DOL of around 0.7 for SNAP-tag labeling with BG-Alexa647[47]. To date, it is not known whether these differences in measured

DOLs are caused by quantification approaches or represent actual differences in achieved DOLs due to sample preparation, cell lines, or the employed substrates.

Protein and peptide tags binding fluorogenic substrates represent an emerging class of labels with potential benefits for counting experiments. Of particular interest are substrates that are quenched in their non-bound state and become brightly fluorescent only upon specific binding to the PP-tag[65–67]. Since non-bound fluorophores are quenched, over-counting due to non-specific background deposition of fluorophores can be significantly reduced.

| **Peptide tags** | | | | | | |
|---|---|---|---|---|---|---|
| | **Tag size** | **Fluorescent label** | **Intra/Extra-cellular (I/E)** | **Covalent label attachment** | **Comments** | **Ref** |
| **FlAsH tag** | 6-10 AA | Bisarsenic fluorophores | I/E | yes | · Brightness of labels is strongly increased upon binding to tag. | [53,68] |
| **SLAP/His tag** | 6-10 AA | Ni-TrisNTA conjugates | (I)/E | no | · Intracellular labeling using trisNTA fusions to cell penetrating peptides | [69,70] |
| **BC2 tag** | 12 AA | bivBC2-Nb | (I)/E | no | · Delivery of non-cell permeable labels via lipid-based transduction for intracellular labeling<br>· Reported DOL for labeling in vitro: 0.64[71] | [71,72] |

| **Protein tags** | | | | | | |
|---|---|---|---|---|---|---|
| | **Tag size** | **Fluorescent label** | **Intra/Extra-cellular (I/E)** | **Covalent label attachment** | **Comments** | **Ref** |
| **SNAP-tag** | 19 kDa | BG conjugates | I/E | yes | · Reported DOLs for labeling on live mammalian cells: 0.43+-0.05 (extracellular tag)[63] and 0.9 (extracellular tag)[73]<br>· Reported DOL after fixation: 0.7[47] | [55] |
| **CLIP-tag** | 19 kDa | BC conjugates | I/E | yes | · Reported DOL for labeling in live mammalian cells: 0.8 (intracellular tag)[73] | [74] |
| **HaloTag** | 30 kDa | HA conjugates | I/E | yes | · Reported DOL for labeling on live mammalian cells: 0.22 +-0.03 (extracellular tag)[63] | [75] |
| **eDHFR** | 18 kDa | TMP conjugates | I/E | (yes) | · Off-target labeling due to label binding to endogeneous DHFR. | [76,77] |

| **Enzyme tags** | | | | | | |
|---|---|---|---|---|---|---|
| | **Tag size** | **Fluorescent label** | **Intra/Extra-cellular (I/E)** | **Covalent label attachment** | **Comments** | **Ref** |
| **Biotin ligase** | 15 AA | Streptavidin conjugate | E | no | · Requires biotin ligase (BirA) protein expression or addition.<br>· Two-step labeling with 1) biotin or biotin derivative 2) Streptavidin-fluorophore conjugate | [78,78] |
| **Lipoic acid ligase** | 12-17 AA | Lipoic acid derivatives | I/E | yes | · One step labeling with LPA-coumarin derivative<br>· Two-step labeling with 1) norbornene derivatives and 2) Tetrazine-TAMRA | [49,79,80] |

**Table 1: Protein and peptide tags for fluorescent labeling of target proteins listed according to tag class.**
Peptide tags consist of a short peptide sequence that is genetically fused to the protein of interest and a corresponding binder which binds the peptide with high affinity. Protein tags consist of a genetically fused protein domain which binds a low molecular weight substrate. Enzyme tags consist of a genetically fused peptide which is recognized by an enzyme that attaches a substrate to the peptide. DOL - degree of labeling. TrisNTA - trivalent N-nitrilotriacetic acid, bivBC2-

Nb - bivariate BC2 nanobody, BG - benzylguanine, BC - benzylcytosine, HA - haloalkane, TMP - trimethoprim. TAMRA - tetramethylrhodamine.

**Labeling with affinity labels**

Fluorophore-conjugated antibodies and other affinity labels represent another class of fluorescent labels widely used in fluorescence microscopy[81,82]. In contrast to fluorescent proteins and PP-tags, affinity labels do not require genetic manipulation of the cell or organism. Instead, affinity labels are raised against a structural motif by inducing an immune reaction in host animals or by in vitro selection. This makes affinity labels a valuable resource for measurements in primary samples where transgene expression is not possible thereby avoiding the problem of under-counting due to unlabeled endogenous proteins. Epitopes targeted by affinity labels typically consist of small, structurally conserved sub-regions of the target molecules. For this reason, a large number of affinity labels directed against many proteins, post-translational modifications, and other cellular components such as lipids or nucleic acids are available[83–85]. As their name suggests, affinity labels bind non-covalently to their target. Their affinity therefore influences labeling efficiency. Also, it has to be kept in mind that labels may dissociate from targets upon prolonged storage. This can be countered by subsequently crosslinking labels with the sample.

Immunoglobulins, like immunoglobulin G (IgG), are widely used for immunofluorescence labeling. However, their large size of ~150 kDa and the fact that each IgG molecule has two epitope binding sites limits their use in quantitative applications. Proteolytic digestion of antibodies into fragments such as F(ab')$_2$ or Fab' with a reduced size of ~48 kDa and ~27 kDa respectively was employed to reduce the size and to obtain single-epitope binders[86]. In contrast, single-chain antibodies such as nanobodies are small in size (~15 kDa), possess only one epitope binding site and can be labeled stoichiometrically using single-cysteine mutants[87–89]. Despite their widespread use as labels in bioimaging, antibodies may exhibit substantial cross-reactivity against non-target molecules. Careful validation of immuno-reagents is therefore key to achieve specific labeling and to avoid off-target label deposition[90,91]. RNA aptamers, DARPins and other protein scaffolds have been proposed as alternative affinity labels with potential uses in labeling for protein copy number determination[92,93]. In contrast to antibodies, these labeling reagents are typically raised *in vitro* and can therefore be produced under strictly controlled conditions without the need to sacrifice host animals. Affinity labels can bind exclusively to a given target such as the actin-binding peptide LifeAct[94], the microtubule binding molecule docetaxel[95] or toxins, like α-bungarotoxin[96]. Aside from adaptable affinity labels that can be raised, these specific labels are in principle suited for molecular counting experiments, given a specific label for the desired target molecule is available.

Quantitative measurements based on affinity labels usually require extensive characterization of the labels and the labeling reaction to determine the number of affinity labels that are bound to one target. For example, the apparent LND will be broader for polyclonal as compared to monoclonal antibodies and other affinity tags. It is obvious that the use of primary and secondary antibodies for indirect immunolabeling further broadens the LND. Since affinities of antibodies were shown to differ depending on the number and type of fluorophores used to label the respective antibodies, such characterizations should always be performed using the same batch of antibodies that is used for subsequent counting experiments[97]. Depending on the fluorescence technique employed for target copy number determination, it may also be required to further characterize the distribution of fluorophores bound to individual affinity labels[19].

Generalized concepts for measuring the labeling efficiency for affinity labels are difficult to establish since affinity labels, by definition, bind directly to their target, for which the exact copy number is yet to be determined. Zanacchi et al. recently introduced a hybrid approach for measuring the degree of labeling of anti-GFP antibodies based on a DNA-origami template[98]. Using GFP-tagged target proteins and the measured DOL, protein copy numbers for different nucleoporins in the nuclear pore complex could then reliably be determined. Instead of determining the absolute degree of labeling, the efficiency of individual labeling protocols can also be investigated by performing titrations with variable affinity label concentrations (fig. 2a)[99]. Such calibrations can then be used to identify optimal label concentrations, but the absolute ratio between labels and target molecules remains unknown.

**Sample preparation**

Efficient and well-defined labeling of target proteins is crucial for enabling in situ protein copy number determination. However, sample preparation – in particular for fixed-cell microscopy – frequently involves additional steps that can have substantial influence on sample morphology and the composition of structures at the molecular level. For instance, most imaging modalities introduced below require static complexes considered being immobile on the time scale of data acquisition. Chemical fixation with aldehydes is a common strategy for effective immobilization of cellular samples, organoids and entire organisms[100–102]. On the downside, most used fixatives, like paraformaldehyde and glutaraldehyde, may cause variable degrees of autofluorescence and thereby alter sample structure and protein localization[103–105]. At the same time, fixation has the potential of epitope degradation adversely affecting post-fixation labeling or of degrading prior-fixation labels, like fluorescent proteins[91]. Alternative immobilization approaches based on reversible or irreversible cryo-arrest have also been suggested for quantitative fluorescence imaging. Here, fixation-induced structural alterations are effectively reduced since the sample is first rapidly frozen and subsequently fixed by addition of chemical fixatives[106–109]. Furthermore, an alternative strategy based on photo-caged glutaraldehyde for rapid chemical fixation with reduced autofluorescent background was recently developed[110]. Permeabilization of cellular membranes is often applied in combination with chemical fixation when labels are not membrane permeable[104,111]. This may additionally alter sample structures and therefore needs to be characterized with respect to its influence on the sample and achievable labeling efficiencies[91]. Overall, labeling turns out to be critical when trying to assess quantitative information of biological structures by use of fluorescence-microscopy-based methods. To convert fluorescence-based estimates into protein numbers not only autofluorescence and unspecific labeling must be considered by additional controls and calibration experiments but also labeling efficiency and stoichiometry as well as potential interference with sample structures.

## Imaging modalities

Conventional far-field light microscopy in the visible to near-infrared spectrum cannot resolve structures below a few hundred nanometers due to diffraction. Since molecular complexes are typically much smaller, individual subunits or binding partners cannot be resolved in space. Although super-resolution methods developed in the past decade are starting to provide near molecular-scale resolution, it is not yet reached in routine experiments. Therefore, dedicated alternative approaches are necessary for quantification of label (or emitter) numbers from fluorescence microscopy experiments. Having discussed important issues of labeling above we now turn towards quantification approaches in conjunction with fluorescence microscopy imaging which are mostly based on single-molecule fluorescence spectroscopy (SMFS).

Among the most prominent SMFS counting methods[112–116] are fluorescence intensity quantification[8], photobleaching step counting[11,117,118], stochastic single-molecule based super-resolution microscopy (SMLM[10,119–121], bSOFI[122] & qPAINT [123]), and quantification based on photon antibunching[124,125] (CoPS).

**Intensity-based counting (IBC)**

A straightforward measure for the number of labels in fluorescence microscopy is the intensity of a complex compared to the unitary fluorescence intensity (or the intensity of a known multimeric standard, fig. 2a)[47,126,127]. Alternatively, the total number of emitted photons can be determined for quantification[128]. For both approaches, quantification relies heavily on the correct measurement of the intensity reference. Conventional dyes and fluorescent proteins can be used provided all labels have equal brightness[129]. This requires homogeneous illumination of the sample or correction for inhomogeneity in post-processing. Aside from the difficulty in creating a high-quality flat illumination field in camera-based imaging, care should be taken for different z-positions especially for TIRF illumination where the illumination intensity strongly decreases with distance from the cover slip. Fluorophore brightness in complex samples is subject to changes in microenvironment e.g. due to solvent effects (viscosity, pH) or fluorescence quenchers[130,131]. For a given average intensity $I$ and corresponding standard deviation $\sigma_I$ of single fluorophores, the intensity of $N$ independent labels is $F_N = N \times I \pm \sqrt{N} \times \sigma_I$ [132]. The increasing uncertainty of molecule number estimates for single complexes limits the method to low numbers. However, analysis of an ensemble of complexes can easily be applied for higher order oligomers or to quantify protein expression in a cell. For example, it was shown that the stoichiometry of FliM proteins in *E. coli's* flagellar motor changes over time[133,134]. Intensity based counting was also used to investigate the stoichiometry and architecture of active DNA replication machinery in *E. coli*[26]. A recent study counted the number of TgDCX and TgAPR1 within organelles of the human parasite *Toxoplasma gondii* in infected fibroblasts[127]. Both proteins are attractive targets for new parasite-specific drugs.

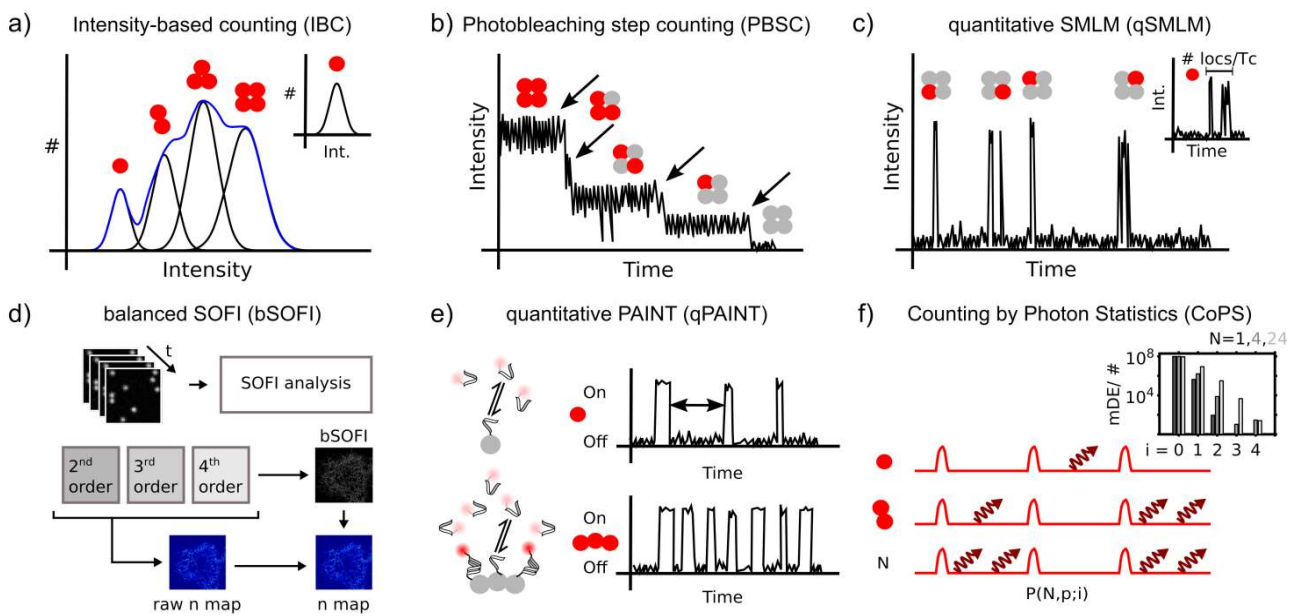

**Figure 2: Quantitative single-molecule-fluorescence-based methods for fluorophore number estimation.** a) Analysis of fluorescence intensity or fluorescence intensity distribution and comparison with an intensity standard (inset), b) Photo-bleaching step analysis, c) Localization microscopy with e.g. calibrated number of localizations per marker or grouping of localizations within short off-times (inset), d) Calculation of molecular density map based on bSOFI analysis up to 4th order, e) Quantitative (q)PAINT analysis of DNA-PAINT off-times, f) Counting by Photon Statistics based on photon antibunching measures the distribution of multiple (coincident) photon detection events (mDE, inset).

**Photobleaching step counting (PBSC)**

A generally accepted hallmark of single molecule observation is a single drop in the intensity trace indicating photo-bleaching. Counting the number of bleaching steps in a complex during continuous illumination thus allows estimation of the number of fluorophores based on photo-destruction (fig. 2b)[135]. Here, the bottleneck lies in the reliable identification of steps in noisy fluorescent transients. The intensity drops become harder to identify for higher $N$ because of accumulating noise (see previous paragraph). As well, the likelihood to miss bleaching events because they occur simultaneously increases exponentially with the label number. Photo-bleaching step counting requires an excellent signal-to-noise ratio and restricts the counting to complexes with a low number of subunits $N$ (typically <10). Data analysis may be carried out automatically, but existing algorithms based on e.g. hidden Markov modeling or change-point analysis apparently only work well for high quality data and often over-fit the number of steps[114,136]. Recently, a Bayesian algorithm was specifically developed for counting large number of photo-bleaching steps (>50), including prior information about dye photo-physics[136]. Here, published datasets were analyzed, e.g. by estimating the number of proteins in RAD51 filaments which play an important role in catalyzing DNA strand exchange and are used in the context of drug delivery.

Instead of trying to resolve individual intensity drops in photo-bleaching traces, one may also correlate the intensity drop and number of fluorescent spots (only decreases once an entire complex is photo-bleached) in a small population over time. This requires only identification of the last photo-bleaching step. The slower the decrease of spot numbers relative to the intensity loss, the higher the number of fluorophores per complex[137]. Counting the number of photo-bleaching steps has been extensively used to quantify membrane receptors and ion channels in the plasma membrane with typical stoichiometries on the order of 4-5[11]. The method is relatively easy to implement given a high-quality single molecule microscope setup, but it inherently destroys the fluorescent labels preventing its use for time-resolved studies of emitter or subunit numbers.

**Quantitative single molecule localization microscopy (qSMLM)**

Usually, diffraction limited microscopy techniques address single complexes individually through spatial separation working at low concentrations. Super-resolution methods facilitate entangling molecular stoichiometries in more crowded environments, albeit requiring additional layers of analysis such as clustering algorithms. Single molecule localization microscopy (SMLM) based super-resolution imaging ((d)STORM, (f)PALM etc.) provides yet another route towards counting fluorescent molecules. Here, the stochastic on-switching of a subset of individual fluorescent molecules (followed by bleaching or off-switching) temporally separates molecules that would otherwise be spatially indistinguishable. The center of individual point-spread functions can be determined with high precision. Merging of all positions of single-molecule localizations over many of such switching/imaging/bleaching cycles yields a final super-resolved image[12]. The number of localizations per complex yields information about the stoichiometry of labeled subunits (fig. 2c). Individual labels may undergo more than one switching cycle and/or exhibit blinking, which results in multiple localizations leading to over-counting. On the other hand, labels that are not activated, localizations that are not registered during data analysis or two labels that are activated simultaneously in a diffraction limited volume give rise to undercounting[48].

One strategy to deal with over-counting is to perform a calibration experiment to extract the number of localizations typically detected per fluorophore or labeled antibody (fig. 3a)[99]. It is crucial to perform calibration experiments in the same cellular micro-environment because otherwise the photo-physics of the fluorophore might change. Other approaches include analysis of the time dependence of blinking and activation[112,138–141], the application of pair-correlation functions[140,141] or Fourier ring analysis[142] (fig 2b). Fitting the distribution of blinking events (on-off-on) of an ensemble of molecules with the probability of the fluorophore to blink as a predetermined parameter is also an option[143–145]. Undercounting can be prevented using appropriate imaging conditions and localization algorithms with a high recall rate[146]. Failure of labels to fluoresce in the imaging channel can be addressed using calibration standards as part of labeling efficiency considerations[48] (section 2).

SMLM-based quantification has the benefit of superior image resolution. It can thus tolerate higher densities of investigated complexes and provide superior contextual information. Yet, it often necessitates the use of clustering algorithms to group localizations originating from the same underlying complex[121]. As the number estimate is based on the blinking kinetics of the fluorophore and thus requires recording of blinking transients, time-resolved measurements will be limited to slow variations of the oligomeric states. The counting range in SMLM quantification appears unrestricted given enough time for imaging of the fluorophores. It is obvious that absolute number determination requires additional efforts to cope with the raised concerns. Both, STORM- and PALM-type localization microscopy have been used for quantification of proteins in recent years. qSMLM has, e.g. shed light on nanoscale organization of membrane proteins involved in cell signaling[147,148], enabled quantification of Bruchpilot molecules in clusters to distinguishes active zone states in Drosophila neuromuscular junctions[99] and allowed tracking of the endosome maturation trajectory in yeast cells[149] (fig, 3a,b).

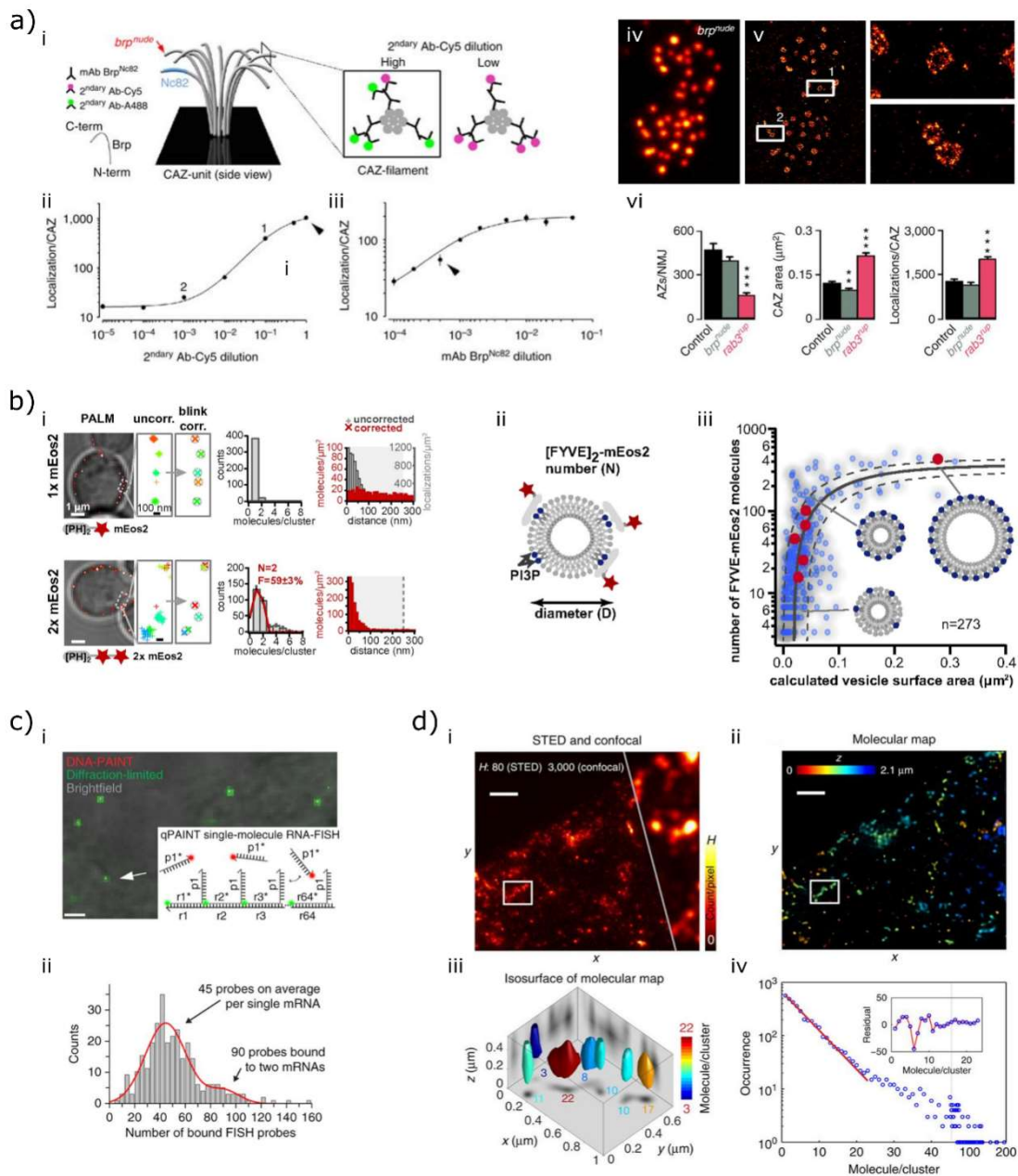

**Figure 3: Selected examples of quantitative number estimation in biological systems.** a) Quantitative dSTORM imaging of Bruchpilot (Brp) in the Drosophila neuromuscular junction (NMJ), adapted from[99]. i) A synaptic active zone cytomatrix (CAZ)-unit (gray) consists of many Brp molecules. The blue shade marks the approximate region of the mAb Brp Nc82 epitope, the red arrow marks the C-terminal (C-term) truncation in brp$_{nude}$. Primary and secondary Ab labeling is characterized by ii) titration of secondary Ab ratio for SMLM keeping the primary and the overall Ab concentration constant to determine the number of localizations per secondary Ab and iii) titration of the primary Ab at constant secondary Ab concentration to provide information on epitope saturation. iv) and v) epifluorescence and dSTORM overview of brp$_{nude}$ spatial distribution in the NMJ. vi) Quantification of imaging data acquired with confocal (left) rank sum test versus controls and localization microscopy (middle and right) rank sum test versus controls for different Brp mutants. b) Quantitative PALM imaging determines the number of accessible phosphatidylinositol 3-phosphate (PI3P) binding sites and the size of individual vesicles in the yeast endocytic pathway, adapted from[149]. i) Superresolution images of calibration constructs with different mEos2 stoichiometry fused to the membrane-localized PH domain of Plc (red) are superimposed on transmitted light images of yeast cells (left); Uncorrected (+) and blinking corrected (X) single molecule positions color-coded by frame number. The number of molecules per cluster is fitted to a binomial distribution (red line) to determine the proportion of red fluorescing mEos2 for multimers F (middle). Pair-correlation functions of corrected images reflect the average distance between molecules, which is constant for the single mEos2 repeat and peaked for the dimer (right). ii) Accessible PI3P binding sites on endocytic/endosomal vesicles are labeled via a tandem repeat of the FYVE domain of EEA1 fused to mEos2. PI3P content and vesicle surface area fall on a characteristic curve; an exponential function (black line) can be fitted to box-smoothed data with 95%

confidence envelope (dashed black line). c) Quantitative PAINT determines the number of smFISH probes bound to SUZ12 mRNA, adapted from[150]. i) 64 smFISH probes with binding sequences unique to a part of the target mRNA (r1*–r64*) carry both a Cy3B label and a single-stranded DNA-PAINT docking strand. ii) qPAINT quantification reveals a ~70% hybridization efficiency of FISH probes (n= 301) with ~45 probes bound to a single mRNA molecule (~90 probes for two mRNAs). d) Mapping the number of transferrin receptors (TfR) in HEK293 cells by measuring photon statistics, adapted from[151]. i) Confocal and STED z-projected images of TfR labeled with anti-TfR aptamer c2. ii) Z-color coded 3D molecular density map generated by photon statistics of both confocal and STED recordings. iii) Isosurfaces rendering of a zoom-in of the molecular map (70% of the overall molecules, box region in i) and ii)). iv) Histogram and exponential fit (red) of the number of molecules in clusters separated by the watershed algorithm in Matlab. Only clusters with up to 24 molecules are considered due to the limited STED resolution not preventing overlap of multiple clusters.

**Balanced super-resolution optical fluctuation imaging (bSOFI)**

SMLM based counting goes hand in hand with increased image resolution. The most common implementations use light driven single-molecule switching. Inappropriate photo-switching rates in combination with high emitter densities can easily lead to artificial clustering[152,153]. Super-resolution optical fluctuation imaging (SOFI) provides an alternative to analyze time series of independent, stochastically blinking fluorophores[154]. It does not require separation of individual emitters but achieves up to n-fold increase in resolution beyond the diffraction limit by computing higher-order statistics (cumulants) across a stack of images[155]. In contrast to localization microscopy, SOFI is compatible with a wide range of blinking conditions, signal-to-noise ratios and high emitter densities[156,157]. More important in this context, the combination of three cumulant orders allow quantitative assessment of molecular parameters (fig. 2d)[158]. Emitter density, brightness and on-time ratio of fluorophores are directly calculated from the image data without the need for calibration measurements for e.g. over-counting corrections. The resolution of the molecular parameter maps is dictated by the lowest cumulant order used in the calculation. Balanced SOFI (bSOFI) enabled determination of the paxilin density in focal adhesions of mouse embryonic fibroblasts[153,157] and was recently used to study the nanoscale distribution and clustering of CD4 glycoprotein mutants in the plasma membrane of T cells. As there are no specific requirements in labeling or calibration, bSOFI can relative easily be used in live-cell experiments. However, like for SMLM, time-resolution is limited by the number of frames required to measure the necessary blinking statistics for a cumulant analysis up to $4^{th}$ order.

**Quantitative point accumulation in nanoscale topology (qPAINT)**

Point accumulation in nanoscale topology (PAINT) is a variant of SMLM that exploits transient binding of dye labeled probes to achieve apparent blinking[159]. It bypasses the need for stochastic photo-switching and is immune to photo-bleaching due to the virtually unlimited source of free-floating probes. DNA-PAINT implements this concept via dye-labeled 'imager' strands that bind to complementary target-bound 'docking' strands[160]. This enables programmable tuning of on- and off-switching kinetics that can be adapted to emitter densities by changing the strand sequence and

'imager' concentration. A drawback of the method is that it is hardly amenable to live-cell imaging. Quantitative PAINT (qPAINT) achieves molecular counting through analysis of the predictable DNA binding kinetics (fig. 2e)[150]. To be precise, qPAINT delivers the number of 'docking' strands, i.e. the number of binding sites and not the number of imaged dyes. DNA hybridization and dissociation can be modeled with a second-order association rate $k_{on}$ and a first-order dissociation rate $k_{off}$. The blinking frequency in the intensity time trace depends on the rate constants, the 'imager' strand concentration and the number of binding sites. First, the mean off-time is determined by fitting the experimentally obtained cumulative distribution function. Then, the number of binding sites is obtained using equation $N = \frac{1}{\xi \tau_{off}}$. This requires prior determination of the imager probe influx rate $\xi = k_{on} \times c_i$ from a known calibration sample under similar conditions. qPAINT decouples counting from photo-physics and enables quantification over a wide dynamic range. It is immune to unaccounted blinking artifacts in SMLM and avoids undercounting due to prematurely photobleached fluorophores. The high precision and accuracy of qPAINT albeit comes at the cost of vastly increased imaging time. qPAINT was first verified by counting the number of individual nucleoporin98 proteins (Nup98) in nuclear pore complexes in U2OS cells and by determining the number of Bruchpilot proteins in the cytomatrix at the synaptic active zone in agreement with earlier studies using dSTORM[99,150]. In the same study, the number of *in situ*–bound smFISH probes per SUZ12 mRNA molecule in fixed HeLa cells was investigated (fig. 3c) [150]. While qPAINT seems to be ideal for quantitative *in vitro* experiments, it appears difficult to extend its use to routine experiments in cell biology. Not only are live-cell applications hindered because of the required labeled DNA strands, but at the same time the long data acquisitions appear to be less attractive as compared to some of the previously described methods.

**Counting by photon statistics (CoPS)**

Another class of counting strategies is based on the fundamental principle of photon antibunching, i.e. the probability for detecting multiple photons vanishes as the detection time window approaches zero[125,161]. Measurements obviously require sensitive detectors for counting single photons such as avalanche photodiodes. Those are commonly used for single molecule imaging in confocal microscopes.

Single fluorophores are isolated quantum systems that emit at most one photon per excitation cycle. The characteristic photon statistics of fluorophores can be used for estimating the number of independently emitting molecules within a diffraction limited spot, i.e. by assuming no photo-physical interaction among different fluorophores. Traditionally, quantification by photon pairs is exploited for counting. The second-order correlation function, i.e., the conditional probability of detecting a photon at time t + τ after the detection of the first photon at time t, can be measured using continuous wave or pulsed laser excitation. The respective measures for antibunching, i.e., the magnitude of the dip in the correlation function or the relative weight of the central peak (coincidence ratio), scale with the number of emitters as $1-\frac{1}{N}$ [162–164]. This function quickly saturates, limiting the reliable counting to the range of 2-3 emitters. Nevertheless, coincidence ratio analysis has been used, e.g. to investigate coupling of emitters in the fluorescent protein tetramer DsRed[165] or to determine the stoichiometry of ion channels[166] and Apolipoprotein A-I in reconstituted high-density lipoproteins[167].

The counting range can be greatly extended if more than two detectors are used, enabling the measurement of photon triples, quadruples etc. Counting by photon statistics (CoPS) uses excitation with short pulses at moderate repetition rate such that maximum one photon is detected per fluorophore and laser pulse. The probability $P_{m,p_b}(N,p;i)$ for i photon detection events then depends on the number of independent emitters N in the focus, on the average photon detection probability per laser pulse and per fluorophore of the microscope setup (short: detection probability p) and on the number of equivalent detectors m[168]. Additional background photons $p_b$ also contribute to the multiple photon detection events. The number of independent emitters N along with their brightness can then be estimated by non-linear regression of the model $P_{m,p_b}(N,p;i)$ to the mDE data (fig. 2f) [169]. In principle counting of large numbers of molecules (at least up to 50 according to simulations with four detectors) is possible. In practice, detector readout electronics limit unbiased estimation to about 20-30 fluorophores due to the high photon count rates[9,151]. The method requires no calibration and provides direct number estimates based on fundamental fluorophore characteristics. For high brightness, counting can be achieved within tens of milliseconds for point measurements, paving the way for quantitative, time-resolved analysis[170]. A variant of CoPS can be used to analyze images acquired by point scanning, incorporating the image formation process in the analysis[151]. Confocal scanning with subsequent STED imaging allows allocating the molecules at higher spatial resolution. Note that the point spread function for i-photon detection or i[th] order antibunching alone supports a $\sqrt{i}$ resolution improvement over one photon detection[171]. Counting by Photon statistics has, e.g. revealed the label number distribution of different types of proteins used for fluorescent tagging in microscopy and bioanalytics[172]. In combination with STED microscopy, the number and three-dimensional nanoscale organization of internalized transferrin receptors in human HEK293 cells were mapped (fig. 3d) [151]]. Thereby it was shown that protein copies can be counted in fluorescently labeled cells without the need for calibration. The relative high time resolution in the order of 100ms for one confocal spot puts also kinetic measurements into perspective. However, CoPS is still limited by the confocal point detection scheme.

## Discussion and outlook

The single molecule counting techniques outlined above vary greatly in their accessible copy number range, their need for reference samples and correction factors as well as the complexity of microscopes, fluorescent labels and data analysis required. As they are all fluorescence-microscopy-based, they are minimally invasive to a controllable extent in respect of the excitation and emission processes and work well in most transparent biological samples. However, all techniques require fluorescence labeling which potentially disturbs biological function, particularly in live-cell experiments. Thus, fluorescence labeling has to be carefully considered as it imposes a strong influence on required control experiments as well as corrections in data analysis. Among the three labeling methods discussed above, conventional fluorescent proteins are well suited for IBC and PBSC, while PALM and bSOFI call for photo-activatable or photo-switchable proteins. CoPS and STORM (and also in many cases bSOFI) are even more demanding as they require brighter and more photo-stable fluorophores which can be introduced by PP-tags or affinity labeling.

However, not only the type of fluorophore itself plays an important role when preparing samples for fluorescence-based protein counting, also the way the label is introduced has a strong influence on the measured label number. Here, PP-tags and FPs are of advantage because they are live-cell compatible and come with an intrinsic 1:1 labeling stoichiometry (if oligomerization has been excluded by genetic engineering) in contrast to affinity labels which have to be applied to fixed cell samples with a usually less well-defined labeling stoichiometry. On the other hand, the latter directly targets endogenous proteins while PP-tags and FPs require genetic modification of the host organism, e.g. by CRISPR/Cas9, to avoid under-counting due to unlabeled endogenous proteins and to achieve physiological expression levels. The latter is considered being critical because over- and under-expression may have a strong influence on oligomerization and hence on biological the function. qPAINT in contrast has thus far been realized with DNA-based probes, i.e. fluorescently labeled DNA-oligonucleotides. Despite their successful use in *in vitro* experiments and in fixed cells, the DNA probes form a severe obstacle for live-cell experiments that could be overcome with the advent of protein or small molecule-based intracellular or cell-permeable probes. Here, the recently developed IRIS probes or genetically encoded affinity labels in combination with fluorogenic dyes are a promising alternative[65,173,174].

It can be concluded that so far none of the three labeling approaches seems to be generally applicable to all quantitative methods discussed here. However, in terms of stoichiometry and live-cell compatibility FPs and PP-tags have clear advantages over affinity labels. When introduced into the organism's genome by gene editing even the problem of unlabeled endogenous protein can be overcome. Furthermore, PP-tags have also the limitation that they require additional steps influencing labeling efficiency: (i) uptake of the fluorescent label into the organism and (ii) subsequent binding of the label to the PP-tag. Finally, ongoing developments to further improve enzyme-mediated attachment of fluorophores to peptide tags might provide an elegant solution to limited labeling efficiencies due to incomplete tag maturation.

The chosen labeling approach also directs the choice of counting standard used for calibration experiments. Figure 4 compares different counting standards with respect to their size and reported numbers of attached fluorophores. Obviously, standards used for calibration measurements should be comparable to the target complexes in both, their physical size and the expected fluorophore number per complex. Other important factors for choosing a standard are robustness, sample compatibility, and the ease of use. Here, the right choice depends on the sample itself. Three different types of calibration standards are shown in figure 4: DNA-based standards, protein oligomers and viruses. DNA-based standards range from small DNA-oligonucleotides (<<10 labels) to well-defined and robust DNA-origami (up to 150 attached labels were reported[129,150]) and can reliably be used for calibration of *in vitro* experiments. Viral particles pose an interesting alternative as they can accommodate even higher fluorophore numbers (>>100 labels) and can be used for calibration of labeling with fluorescent proteins or PP-tags[127]. Similar to DNA probes, viral capsids are not suited as standards in living cells. Protein oligomers, on the other hand, are ideally suited standards for *in situ* and live-cell experiments because they can be transiently or stably expressed in living cells[47,135,175,176]. They are quite versatile in terms of fluorescent labels as the monomeric subunits can be fused to  they would require permeabliziation and incubation.

On the side of microscopic imaging, additional problems with protein oligomers expressed for calibration in cells might arise from the heterogeneous spatial distribution of the complexes leading to intensity variations and cross-talk among different structures due to overlapping point-spread functions. Here, membrane-localized complexes such as plasma-membrane receptor oligomers or nuclear pore complexes have the advantage of localizing in a membrane that can be placed in the imaging plane, thereby yielding relatively homogeneous intensities[135,177]. The strength of such effects can vary with protein expression levels, cell thickness and shape making it dependent on cell type.

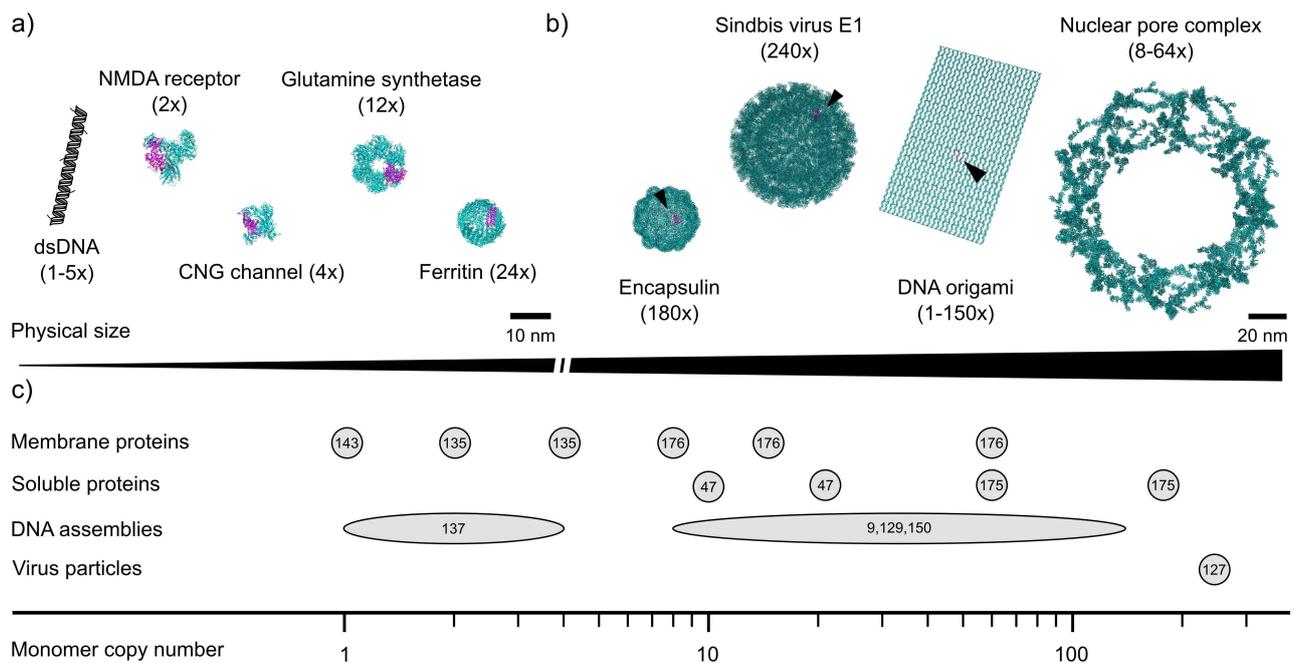

**Figure 4: Standards for protein copy number measurements.** Protein oligomers and DNA-based probes allow to attach a defined number of fluorophores to complexes with variable size. a) Short DNA oligonucleotides can be covalently labeled with organic fluorophores. Different trans-membrane and soluble protein complexes with monomer copy numbers ranging from 2-24 have been described in the literature. b) Larger protein complexes and DNA origami can be used to obtain standards with up to 240 fluorophores within a structure <100 nm. Note the difference in scale for structures in a) and b). c) different classes of molecules with a variable number of fluorophores per structure have been described in the literature (refs indicated by numbers in circles). Protein structures in b) were rendered using Chimera[7] based on publicly available data deposited in PDBe. Accession numbers: 5iou (NMDAR), 5h3o (CNG), 1fpy (Glutamine synthetase), 1eum (Ferritin), 4pt2 (Encapsulin), 3j0f (Sindbis virus), 5a9q (Nuclear pore complex). The structure of a rectangular DNA origami[178] was predicted using CanDo [179,180] and rendered using Chimera.

In terms of the counting techniques, the achievable counting range is very important showing large variations among the different methods (table 2). A typical application for photo-bleaching step analysis is the analysis of membrane receptors or ion channels in the counting range of 1-5. Intensity reference measurements can determine much higher average numbers (e.g. protein copy numbers in a cell) but lose the ability to quantify single probe molecules due to high variations in intensity. Localization microscopy and qPAINT based quantification is also not restricted to small numbers and was experimentally shown to count to more than 33 and 150 labels respectively. However, qSMLM requires significantly more effort to determine absolute numbers and the application range may depend on the calibration method chosen. The absolute counting range of bSOFI has so far not been determined as it has only been used for relative number estimation not considering effects of labeling with a focus on protein clustering[153,157]. For CoPS it was shown that up to 30 labels can be reliably counted[9], a significant extension with respect to early antibunching based quantification. So far, bias and precision has not been determined for all the methods which together with the variations in the counting range complicates direct comparisons. For future developments it will therefore be important to find or design standardized targets enabling benchmarking of existing and newly developed quantitative methods.

To keep experiments as simple as possible additional control experiments should be kept at minimum. Consequently, quantitative microscopy should return absolute copy numbers which is so far only achieved by PBSC and CoPS. The other methods require calibration and comparison with defined standards. As the underlying parameters for quantification, i.e. brightness for IBC and kinetics for qSMLM, bSOFI and qPAINT, may depend on the local chemical environment, control experiments will always have to be carried out in the sample under study. Additionally, general methods to determine the label efficiency are required independent of the counting method, since in this context a reproducible and known labeling efficiency is essential.

Biological function is not only reflected in cellular structures and their protein stoichiometry but furthermore in the dynamics of how these structures are formed, reshaped and disassembled. Therefore, functional understanding of the underlying kinetics calls for time-resolved measurements of protein numbers. In general this requires rapid and non-destructive data acquisition which in combination with imaging can relatively easily be achieved by IBC. In contrast to the other methods, PBSC is *per se* destructive as it is based on photo-bleaching and thus conceptually not suited for time-resolved measurements. qSMLM and qPAINT require relative long acquisition times for accumulating enough frames for image reconstruction. Time-resolved counting is therefore limited to relatively slow processes on timescales >10 s. Currently, this is also the case for CoPS where raster scanning is required for imaging thereby limiting the time resolution to >1s or more depending on the field of view[151]. This could, however, be drastically decreased by the development of novel, single-photon sensitive imaging detectors. Here the 100-200 ms required for a valid CoPS estimate in point measurements set a lower limit for the speed of the image acquisition[170]. An imaging single-photon detector with sufficiently high frame-rates and quantum yield would at the same time increase throughput of the technique which is important for achieving a robust statistical measurement. Further technological progress in single-photon imaging thus bears the potential of extending quantitative imaging techniques to shorter time scales.

For applications in cell biology also imaging in three dimensions and improved resolution has become key. Thus, it is important to discuss to which extent these criteria can be met by the presented methods. Improved resolution is certainly of advantage in combination with a quantitative approach especially when the density of the structures is too high to be resolved by diffraction-limited techniques. So far, resolution improvements are intrinsically achieved by qSMLM, bSOFI, and qPAINT while other methods provide information limited by diffraction. For CoPS improved resolution has been shown by using the measured photon-statistics[181] or by combination with STED microscopy[151]. But combination with external super-resolution techniques requires more extensive experiments and increased photon-influx complicating high-throughput live-cell studies.

Like for resolution improvement, quantitative 3D imaging is of advantage if the structures under study come at high spatial density or even necessary when the protein of interest has a 3D distribution in the cell. Most of the underlying imaging modalities of the quantitative methods are amenable and have readily been implemented with established 3D approaches, like astigmatism and dual plane imaging for SMLM[182–184], multiplane imaging for 3D SOFI[185] or perpendicular illumination using slight sheet approaches[186,187]. However, and to the best of our knowledge, protein counting in 3D structures has so far not been demonstrated with any of the imaging modalities, possibly due to imaging side-effects, like sparse blinking, out-of-focus-bleaching and increased background. For similar reasons implementation of 3D will be difficult to achieve for PBC. However, photon statistics modeling with CoPS has been extended to 3D[151], albeit confocal point scanning leads to slow imaging. An alternative approach to quantitative 3D imaging avoiding the aforementioned side-effects is physical sectioning of resin-embedded samples, i.e. array tomography[188–190]. Although only fixed samples can be studied it would offer the advantage of enabling the combination with, for instance, electron microscopy as correlative approach. Clearly, all the mentioned possibilities will require a high level of automation and computational resources for data acquisition, storage and processing.

From the previous discussion, it is not yet clear which of the different methods has the greatest potential for becoming the most universal quantitative microscopy approach. Certainly, a detailed benchmarking will be required in the future to clarify the potentials and limitations of the methods described above. It is also obvious that some of the methods have not yet been developed to their full potential. CoPS could benefit from sophisticated hardware regarding 3D imaging, spatial resolution and throughput[191–193]. Similarly, more progress can be expected for bSOFI. Here, it will be important to see if the acquired density maps suited for relative comparison can be linked to absolute numbers, e.g. by means of using a counting standard. Future developments in qPAINT will strongly depend on novel probes with similar, advantageous properties as DNA-oligonucleotides paving way to live-cell experiments. Here, affinity-based peptide labeling has a chance to fulfil promises of reversible binding and multiplexing based on binding affinity.

Overall, quantitative microscopy methods will only be routinely used if they are robust, reliable, and easy to use. Whether or not absolute numbers are measured will be of less importance if calibration for this method is simple and fast. For calibration, the minimum requirement is a method to measure label efficiency. Independent of the technique, the ambitious goal of extracting reliable protein numbers can only be reached in a combined approach of developing the microscopy technique along with corresponding sample preparation and labeling approaches as well as robust calibration experiments. This requires close collaboration of experts in different areas ranging from optics, over biochemistry, molecular biology and cell biology to data processing. In the future, microscopy in biology and potentially in medical research will require a high level of automation to enable screening of different structures under systematic variation of conditions and effectors. Additionally, it would be advantageous to acquire protein copy numbers along with structural information to allow for correlative approaches and simultaneous experimental validation of multiple parameters to be compared with quantitative biological models. This demands for universal pipelines among different integrated microscopy techniques (see Sibarita et al. for first steps in that direction[194]). Universality in this context clearly involves improved storage and processing capacities for large amounts of data. Taken together, developments in quantitative microscopy do not only require great progress of individual techniques but beyond this a high level of standardization to enable robust comparison of similar methods and their integration with existing techniques.

|  | Counting range established via simulations or measurements with calibration standards | Pitfalls for live-cell imaging | Time-resolution |
|---|---|---|---|
| **IBC** | · 1-36 experiments with DNA origami[129]<br>· 10-24 protein oligomer standards[47]<br>· 240 virus particles[127]<br>· 100s – 10.000 total protein per cell* [195] | · none | > 0.1 sec (detector limited) |
| **PBSC** | · 1-5 experiments with membrane receptors[11]<br>· >50 simulations with advanced algorithm[136] | · phototoxicity | n/a (photodestruction) |
| **qSMLM** | · 1-3 experiments membrane protein standards[143]<br>· 33 experiments on FliM flagellar motor[112]<br>· 1-200 (or more) simulations[112] | · phototoxicity esp. UV light | > 10 sec |
| **bSOFI** | · no validation with calibration standard yet<br>· only relative density reported in experiments so far<br>· 100-1600 molecules/µm$^2$ simulations[196] | · phototoxicity esp. UV light, but lower light dose than qSMLM | > 10 sec |
| **qPAINT** | · 1-500 simulations[150]<br>· 1-150 experiments with DNA origami[150] | · DNA-based probes not cell permeable<br>· potentially achievable with genetically encoded PAINT probes | > minutes - hours |
| **CoPS** | · 1-50 simulations[168]<br>· 20-30 experiments with DNA origami[9] | · phototoxicity in combination with STED<br>· so far, no FP compatibility shown | << 1 sec (point-based)<br>> 1-10 sec (imaging) |

**Table 2: Counting ranges, compatibility with live-cell imaging and time resolution are key parameters for comparing imaging modalities.** n/a: not applicable. *Measurements were cross-validated by immunoblotting.

## Acknowledgements

We gratefully acknowledge the German Research Foundation (DFG, HE 4559/6-1) and the Federal Ministry for Education and Science (BMBF/VDI, MorphiQuant-3D) for their financial support. K.S.G. has received funding from the European Union's Horizon 2020 research and innovation program under the Marie Skłodowska-Curie Grant Agreement No. [750528]. We also thank F. Braun and J. Hummert for fruitful discussions and A. Radenovic for supporting K.S.G. in writing this perspective.